\documentclass[aps,twocolumn,showpacs,floatfix]{revtex4}
\pdfoutput=1
\usepackage{epsfig,here,graphicx}
\usepackage{amsmath,amssymb,amsfonts,color}

\newcommand{\rem}[1]{}

\newcommand{\ui}{\mathrm{i}}

\newcommand{\ud}{\mathrm{d}}

\renewcommand{\Im}{\operatorname{Im}}

\begin{document}

\title{Unidirectional Emission from Circular Dielectric Microresonators
with a Point Scatterer}

\author{C. P. Dettmann$^{1}$, G. V. Morozov$^{1}$, M. Sieber$^{1}$,
H. Waalkens$^{1,2}$}
\affiliation{$^{1}$Department of Mathematics, University of Bristol,
Bristol BS8 1TW, United Kingdom\\
$^{2}$Department of Mathematics, University of Groningen, 9747 AG Groningen,
The Netherlands}

\date{\today}

\begin{abstract}
Circular microresonators are micron sized dielectric disks embedded in material
of lower refractive index. They possess modes of extremely high Q-factors
(low lasing thresholds) which makes them ideal candidates for the realization of
miniature laser sources. They have, however, the disadvantage of isotropic
light emission caused by the rotational symmetry of the system. In order to obtain
high directivity of the emission while retaining high Q-factors, we consider
a microdisk with a
pointlike scatterer placed off-center inside of the disk. We calculate the resulting
resonant modes and show that some of them possess both of the desired characteristics.
The emission is predominantly in the direction opposite to the scatterer.
We show that classical ray optics is a useful guide to optimizing the design parameters
of this system. We further find that exceptional points in the resonance spectrum
influence how complex resonance wavenumbers change if system parameters are varied.
\end{abstract}


\maketitle

%%%%%%%%%%%%%%%%%%%%%%%%%%%%%%%%%%%%%%%%%%%%%%%%%%%%%%%%%%%%%%%%%%%
%%%%%%%%%%%%%%%%%%%%%%%%%%%%%%%%%%%%%%%%%%%%%%%%%%%%%%%%%%%%%%%%%%%

\section{Introduction} 

Thin dielectric microcavities of different shapes are widely used as microresonators 
in laser physics and integrated optics
(see \cite{Vahala2003,Ilchenko2006} and references therein).
The new directions in  microcavity laser research
have been recently reviewed in \cite{NosichSmotrovaBoriskinaBensonSewell2007}.
Microresonators are open systems coupled to the external world,
and therefore do not have bound states. 
Instead, their eigenmodes (resonances) 
are characterized by complex wavenumbers $k=k_r + i k_i = \omega/c -i/(c\tau)$.
Here $\tau$ is the lifetime of the resonance 
and $c$ denotes the speed of light. 
The so-called resonance quality factor ($Q$ factor for short)
is then defined as $Q = k_r/2\,|k_i|$, and is a measure
of the suitability of a mode for lasing.

The resonances of any passive (no active lasing particles) microcavity 
filled with nonmagnetic (${\bf B}={\bf H}$) dielectric material 
are the solutions of the time-independent Maxwell equations
\begin{equation}
\nabla  \times {\bf E} = ik\,{\bf H}, \quad \nabla  \times {\bf H} = 
- ik\,\varepsilon \left( {\bf r} \right){\bf E}
\end{equation}
when appropriate boundary conditions on the fields $H$ and $E$ are imposed.
When the fields are independent of $z$, which
is strictly speaking only the case for a cylindrical microcavity of infinite
length (see below), 
the above equations in cylindrical coordinates ($r,\varphi,z$) reduce to
\begin{equation}
\begin{split}
\frac{1}{r}\frac{{\partial E_z }}{{\partial \varphi }} = ik\,H_r,  \quad 
- \frac{{\partial E_z }}{{\partial r}} = ik\,H_\varphi,   \\ 
\frac{1}{r}\left[ {\frac{{\partial \left( {rE_\varphi  } \right)}}{{\partial r}}
- \frac{{\partial E_r }}{{\partial \varphi }}} \right] 
= ik\,H_z, 
\end{split}
\end{equation}
and
\begin{equation}
\begin{split}
\frac{1}{r}\frac{{\partial H_z }}{{\partial \varphi }} = -ik\,\varepsilon \left(r,\varphi \right)E_r,\,\,\,\,
- \frac{{\partial H_z }}{{\partial r}} = -ik\,\varepsilon \left(r,\varphi \right)E_\varphi,   \\ 
\frac{1}{r}\left[ {\frac{{\partial \left( {rH_\varphi  } \right)}}{{\partial r}} -
\frac{{\partial H_r }}{{\partial \varphi }}} \right] = -ik\,\varepsilon \left(r,\varphi \right)E_z. 
\end{split}
\end{equation}
The solutions can be separated into TM waves 'transverse magnetic, i.e.  $H_z=0$)
and TE waves `transverse electric', i.e. $E_z=0$).
Introducing the position dependent refractive index $n(r,\varphi) \equiv \sqrt{\varepsilon(r,\varphi)}$, 
we get for TM modes a scalar wave equation for $E_z$,
\begin{equation}
\displaystyle \frac{{\partial ^2 E_z }}{{\partial r ^2 }} + \frac{1}{r }\frac{{\partial E_z }}{{\partial r }} +
\frac{1}{{r ^2 }}\frac{{\partial ^2 E_z }}{{\partial \varphi ^2 }} + k^2 n^2 \left( {r ,\varphi } \right)E_z  =  0\,.
\end{equation}
The magnetic field $H$ is then recovered from
\begin{equation}
H_r  =  - \displaystyle\frac{i}{{kr}}\frac{{\partial E_z }}{{\partial \varphi }}, \quad
H_\varphi  =  \frac{i}{{k\,}}\frac{{\partial E_z }}{{\partial r }}\,.
\end{equation}
For TE modes,  it is advantageous to introduce the new field 
$h_z(r,\varphi) \equiv H_z(r,\varphi)/n(r,\varphi)$.
This field satisfies again a scalar wave equaton,
\begin{eqnarray}
\frac{{\partial ^2 h_z }}{{\partial r^2 }}  + \frac{1}{r}\frac{{\partial h_z }}{{\partial r}}
+ \frac{1}{{r^2 }}\frac{{\partial ^2 h_z }}{{\partial \varphi ^2 }} + k^2 n^2(r,\varphi)\,
h_z (r,\varphi) \nonumber \\  + \left[\frac{\nabla^2 n}{n(r,\varphi)} - 2\left( 
\frac{\nabla n}{n(r,\varphi)} \right)^2 \right]  h_z (r,\varphi) = 0\,,
\end{eqnarray}
and the electric field $E$ can be recovered in this case from
\begin{equation}
\begin{split}
E_r   = \frac{i}{{krn(r,\varphi)}}\left[ {\frac{{\partial h_z }}{{\partial \varphi}}
+ \frac{1}{{\,n(r,\varphi)}}\frac{{\partial n}}{{\partial \varphi}}\,h_z(r,\varphi)} \right], \\
E_\varphi   =  - \frac{i}{{k\,n(r,\varphi)}}\left[ {\frac{{\partial h_z }}{{\partial r}} + 
\frac{1}{{\,n(r,\varphi)}}\frac{{\partial n}}{{\partial r}}\,h_z(r,\varphi)} \right].
\end{split}
\end{equation}
The applicability of these equations can be extended from
infinitely long cylindrical  microcavities to flat microcavities  
for which the cavity thickness is only a fraction of the mode wavelength.
The refractive index has then to be replaced by an effective refractive
index $n_{{\rm eff}}(r,\varphi)$
which takes into account the material as well as the thickness of the cavity. 
We note that it is in general a difficult problem to explicitly  compute $n_{{\rm eff}}(r,\varphi)$
for a given material and thickness (see, e.g., Ref.~\cite{Bogomolny2007}), and one therefore 
typically resorts to using an experimentally determined effective refractive index instead.
 
The simplest microcavity shape is a thin circular microdisk.
The technological progress in recent years has made possible
the construction of microdisks in the $\mu$m-domain. 
They are natural candidates for the construction of lasers
since some of their modes have extremely high $Q$-factor (low thresholds)
\cite{LeviSlusher1992,LeviSlusher1993}.  
In those modes, which are called `whispering gallery modes', 
light is trapped by total internal reflection and  circulates along the circumference of the disk.
As a consequence of the rotational symmetry the light emission of microdisks is isotropic. 
For many applications, however, a directional light emission is required.
%more precisely the emission displays $2m$ identical beams,
%where $m =0,1,2 ...$, see the next section. 
To obtain a directional emission we recently proposed to break the symmetry of a microdisk by placing a point scatterer
inside but not at the center of the microdisk (see Ref.~\cite{DettmannMorozovSieberWaalkens2008}). 
We have demonstrated that such a geometry leads to a significant enhancement
of the directivity
of some TM modes in outgoing light while preserving their high $Q$-factors.
Other attempts to breaking the symmetry include the introduction of some other
defects  inside the microdisk 
like a linear defect \cite{ApalkovRaikh2004,TulekVardeny2007} 
or a hole \cite{WiersigHentschel2006}. Another approach is to deform the
boundary of the cavity
\cite{LeviSlusherGlass1993,NockelStoneChang1996,Lebental2006,LeeKim2007}, or 
to couple light into and out of a microdisk with the aid of an optical fiber
taper waveguide \cite{SrinivasanPainter2007}. 
However, the advantage of our method
is the analytic tractability which allows for a systematic 
optimization of the design parameters (location and strength of the scatterer)
with only modest numerical efforts.

The purpose of this paper is to extend the theory in \cite{DettmannMorozovSieberWaalkens2008}  to TE modes, 
and give a systematic study of the appearance of both highly directional 
TM and TE modes and its dependence on the distance of a point scatterer from the disk center.
In particular, we  provide arguments based on geometric optics to explain this dependence.
This paper is organized as follows.
In the following section, Sec.~II, we use a Green's function method
to calculate the positions of the resonant modes of a microdisk 
with a point scatterer in the complex wavenumber plane. 
In Sec.~III we discuss in some detail the physical interpretation of the 
parameter that describes the strength of the point scatterer. In Sec.~IV we investigate
in detail the directivity of the resonance modes for the 
microdisk with point scatterer, and in Sec.~V we show that classical
ray optics is a useful guide to optimizing the design parameters
of the system. In Sec.~VI we investigate the role of exceptional
points in our system, and we finish with some concluding remarks in Sec.~VII.

%%%%%%%%%%%%%%%%%%%%%%%%%%%%%%%%%%%%%%%%%%%

\section{Theory of Microdisk Cavities with a Point Scatterer} 

Let $\Psi$ stand for $E_z$ in the case of TM polarization
and for $h_z$ in the case of TE polarization.
For a homogeneous dielectric microdisk of radius $R$
and effective refractive index $n$ in a medium
of refractive index $1$, Eqs. (4) and (6) take the form
\begin{equation}
\label{eq:Psi}
 \frac{{\partial ^2 \Psi }}{{\partial r ^2 }} + 
\frac{1}{r }\frac{{\partial \Psi }}{{\partial r }} + 
\frac{1}{{r ^2 }}\frac{{\partial ^2 \Psi}}{{\partial \varphi ^2 }} +
k^2 n^2 \Psi \left( {r ,\varphi } \right) = 0,
\end{equation}
inside the microdisk ($r<R$) and the same form with $n$ replaced by $1$ 
outside the microdisk ($r>R$).
The resonances  are obtained by imposing outgoing boundary conditions
at infinity, i.e. we require that 
$\Psi \left( r \right) \propto {{e^{ikr} } \mathord{\left/
 {\vphantom {{e^{ikr} } {\sqrt r }}} \right.
 \kern-\nulldelimiterspace} {\sqrt r }}$,
$r \rightarrow \infty$.
Moreover, for physical reasons the value of the EM field at the disk center 
must be finite.
At the boundary of the microdisk, $r=R$,
the electric field component $E_z$ and its derivative 
has to be continuous for TM modes. 
Similarly,  for TE modes, the field $h_z$ multiplied by the refractive index 
and its derivative divided by the refractive index has to be continuous at $r=R$.
These boundary conditions lead to the 
`whispering gallery' (WG) modes
\begin{equation} \label{modes}
\Psi ^m  = \left\{ {\begin{array}{*{20}c}
   {N_m J_m \left( {knr} \right)\left({\begin{array}{*{20}c}
   {\cos m\varphi }  \\
   {\sin m\varphi }  \\
\end{array}} \right), \quad r < R,}  \vspace{8pt}\\
   {H_m \left( {kr} \right)\left( {\begin{array}{*{20}c}
   {\cos m\varphi }  \\
   {\sin m\varphi }  \\
\end{array}} \right), \quad r > R,}  \\
\end{array}} \right.
\end{equation}
where for TM modes, the complex wavenumbers $k$ are solutions of
\begin{equation} 
\label{eq:unpert_quant_cond_tm}
A_{m}^{\rm TM} \equiv J_m( k n R )H_m^{\,\,\,'} (k R) 
- n\,J_m^{\,\,\,'} ( k n R )H_m ( k R ) = 0,
\end{equation}
and for TE modes, the complex wavenumbers $k$ are solutions of
\begin{equation} 
\label{eq:unpert_quant_cond_te}
A_{m}^{\rm TE} \equiv n\,J_m( k n R )H_m^{\,\,\,'} (k R) 
-J_m^{\,\,\,'} ( k n R )H_m ( k R ) = 0.
\end{equation}
Here $J_m$ and $H_m$ are Bessel and Hankel functions of the first kind respectively, 
$m=0, 1, 2, ...$ is the azimuthal modal index.
The constants $N_m$ are given by 
\begin{equation}
N_m = 
\begin{cases}
H_m(kR)/J_m(knR)     & \text{for TM modes} \\
H_m(kR)/(n J_m(nkR)) & \text{for TE modes}\,.
\end{cases}
\end{equation}
Physically, the azimuthal modal index $m=0,1,2,\ldots$ characterizes 
the field variation along the disk circumference, 
with the number of intensity hotspots being equal to $2m$.
The wavenumbers, $k$, are twofold degenerate for $m>0$, and nondegenerate for $m=0$.  
The radial modal index $q=1, 2, ...$ will be used to label different resonances
with the same azimuthal modal index $m$.  
For resonances which are relatively close to the real axis of the the complex wavenumber plane (so called `internal' or `Feshbach' resonances), 
the index $q$ gives in general the number of intensity spots in the radial direction. As discussed in detail in \cite{DettmannMorozovSieberWaalkens2009} there are exceptions to this rule for some TE internal resonances.

We note that for each fixed $m$, there exist further solutions 
of Eqs.~(\ref{eq:unpert_quant_cond_tm},\ref{eq:unpert_quant_cond_te})  (so called `external' or `shape' resonances)
which are typically located deeper in the lower half of the complex  wavenumber planes compared to internal resonances (see Refs.~\cite{Bogomolny2008,Lee2008}). 
The external resonances are very leaky
(low $Q$-factors) and, as a result, can not be directly used for lasing. However,
they are of theoretical interest in their own right. 
To properly distinguish between external and internal resonances  
one needs to trace the resonances as a function of the refractive index $n$ to the values they obtain in the limit
$n \rightarrow \infty$. As discussed in detail in \cite{DettmannMorozovSieberWaalkens2009}
this limiting value is real for internal resonances, and complex (not real) for external resonances. 
The procedure of tracing the resonances in the complex wavenumber plane is especially important for TE modes for which some of the external
resonances lie in the same domain of the complex wavenumber plane 
as the internal resonances. For  an example of this phenomenon, we refer to  Sec.~IV. 

In \cite{DettmannMorozovSieberWaalkens2008} we derived the 
Green's function for the TM modes of a microdisk. Following  
 Ref.~\cite{MorseFeshbach1953} it is given by
a sum over all angular harmonics $e^{\ui m(\varphi-\varphi_0)}$
multiplied by the corresponding radial parts. Using the same method it is not difficult
to derive the Green's function also for TE waves. In fact, both functions can be written as
\begin{widetext}
\begin{equation} 
\label{eq:G_is}
G^{\text{TM/TE}}\left( {{\bf r},{\bf r}_0, k} \right) = \left\{ {\begin{array}{*{20}c}
   { - \displaystyle\frac{\ui}{4}H_0 \left( {kn \left| {{\bf r} - {\bf r}_0 }
   \right|} \right) + \displaystyle\frac{\ui}{4}\displaystyle\sum\limits_{m  =
   0}^\infty  {\displaystyle\frac{{C^{\text{TM/TE}}_{m} }}{{A^{\text{TM/TE}}_{m} }} \epsilon_m
   \cos\left[ m\left( {\varphi  - \varphi _0 } \right)\right]J_m \left( {kn r_<  }
   \right)J_m \left( {kn r_ >  } \right)} ,} & {r_ <  ,r_ >  \, < \,R,}  \\
   {  \displaystyle\frac{1}{2\pi k R}\displaystyle\sum\limits_{m = 0}^\infty 
   {\displaystyle\frac{1}{{A^{\text{TM/TE}}_{m} }}\epsilon_m\cos \left[m\left( {\varphi  - \varphi _0 }
   \right)\right]J_m \left( {kn r_ <  } \right)H_m \left( {k r_ >  } \right)} ,}
   & {r_< < R < r_> ,}  \\
   { - \displaystyle\frac{\ui}{4}H_0 \left( {k \left| {{\bf r} - {\bf r}_0 }
   \right|} \right) + \displaystyle\frac{\ui}{4}\displaystyle\sum\limits_{m  =
   0}^\infty  {\displaystyle\frac{{B^{\text{TM/TE}}_{m} }}{{A^{\text{TM/TE}}_{m} }} \epsilon_m \cos\left[ 
   m\left( {\varphi  - \varphi _0 } \right)\right]H_m \left({kr_<} \right)
   H_m  \left( {kr_>} \right)} ,} & {r_ <  ,r_ >  \, > \,R,}  \\
\end{array}} \right.
\end{equation}
\end{widetext}
where $r_{<}$ ($r_{>}$) is the smaller (larger) of $r$ and $r_0$.
The coefficients are $\epsilon_m=2$ if $m\ne0$, $\epsilon_m=1$ if $m=0$, and
\begin{equation} \nonumber
\begin{split}
B_{m}^{\rm TM} & =  J_m \left( {knR} \right)J_m^{\,\,\,'} \left({kR} \right)
 - n\,J_m^{\,\,\,'} \left( {kn R} \right)J_m \left( {kR} \right),  \\ 
C_{m}^{\rm TM}& =  H_m \left( {kn R} \right) H_m^{\,\,\,'} \left( {k R} \right)
 - n\,H_m^{\,\,\,'} \left( {kn R} \right)H_m \left( {kR} \right), \\
B_{m}^{\rm TE} & =  n\,J_m \left( {knR} \right)J_m^{\,\,\,'} \left({kR} \right)
 - J_m^{\,\,\,'} \left( {kn R} \right)J_m \left( {kR} \right),  \\ 
C_{m}^{\rm TE}& =  n\,H_m \left( {kn R} \right) H_m^{\,\,\,'} \left( {k R} \right)
 - H_m^{\,\,\,'} \left( {kn R} \right)H_m \left( {kR} \right). 
\end{split}
\end{equation}
The resonances of the microdisk are then
determined by the poles of the Green's function i.e. by $A^{\text{TM/TE}}_m=0$.
This agrees with  resonance conditions (\ref{eq:unpert_quant_cond_tm}) and
(\ref{eq:unpert_quant_cond_te}).
We note that the resonance wavefunctions
are exponentially increasing as $r \rightarrow \infty$, and hence cannot be
normalized. However any constant
multiplying the resonance wavefunctions can be fixed by comparing the wavefunctions
to the residues of the Green's function.

Using methods of self-adjoint extension theory \cite{Zorbas1980,Shigehara1994},
we showed in Ref.~\cite{DettmannMorozovSieberWaalkens2008} that the presence of a
point scatterer, which is located at a point ${\bf d}$ on the $x$-axis ($\varphi=0$),
leaves the resonances of the unperturbed disk (WG modes) with the angular part
$\sin(m\varphi)$ unchanged, while the complex wavenumbers $k_{\rm res}$ of the resonances
with the angular part $\cos(m\varphi)$  become solutions of the transcendental equation
\begin{equation} \label{rescond}
1 - \lambda G_{\mathrm{reg}}({\bf d},{\bf d}, k) = 0 \, .
\end{equation}
Here $G_{\mathrm{reg}}$ is the regularized Green's function which is obtained by subtracting
the logarithmically  divergent term $\ln( k_0 |{\bf r} - {\bf r_0}|)/ 2 \pi$ from the Green's
function (\ref{eq:G_is}) in the limit ${\bf r}$, ${\bf r_0} \rightarrow {\bf d}$. 
The two parameters $\lambda$ and $k_0$ can be absorbed in a single  new parameter $a$ defined by
$2 \pi \equiv - \lambda \ln k_0 a$. Then the condition for resonances (\ref{rescond}) becomes
\begin{equation}
\label{eq:quant_cond}
\begin{split}
0 = &  -\frac{\ui}{4}+\frac{{1}}{2\pi }\big( {\ln \frac{{k_{\,\rm {res}} n a }}{{2 }} + \gamma } \big)\\
& + \frac{\ui}{4}\sum_{m=0}^\infty {\frac{C_m(k_{\rm res})}{A_m(k_{\rm res})}}\,
\epsilon_m {J^2_m \left( {k_{\,\rm {res}}nd } \right)}\,,
\end{split}
\end{equation}
where  $\gamma \approx 0.5772$ is the Euler-Mascheroni constant.
The parameter $a$ then determines  the ``strength'' of the point scatterer
located at the distance $d = |{\bf d}| <R$ from the center. 

The corresponding resonance wavefunction is given by
\begin{equation} \label{wave}
\Psi({\bf r}) = N G({\bf r},{\bf d},k_{\rm res})\,,
\end{equation}
where $G$ is the Green's function in Eq.~(\ref{eq:G_is}), and $N$ is a normalization factor. 
Outside of the microdisk, i.e. in the region $r>R$, 
the field takes the form 
\begin{equation}
\label{eq:outside_field}
\Psi  = \frac{N}{{2\pi k_{\rm res}R}}\sum\limits_{m = 0}^\infty
{\frac{{\epsilon _m\cos\left( m {\varphi} \right) J_m \left(
{k_{\rm res}nd} \right) }}{{A_m }}}H_m \left( {k_{\rm res}r} \right).
\end{equation}
In principle, the normalization factor can be obtained again from the
residue of the Green's function of the perturbed system (see Eq.~(\ref{greenperturb}) in the next section).

From Eq.~(\ref{eq:quant_cond}) we see that in the limit $a=0$ and $a\rightarrow\infty$ we recover the resonances of the unperturbed microdisk (without a point scatterer). 
For $a$ close to 0 or infinity, the approximate solutions of 
Eq. (\ref{eq:quant_cond}) can be found perturbatively.
In leading order one finds
\begin{equation}
\label{eq:approx_quant_cond}
k_{{\rm res}}  \cong k - {\ui}\,\pi \frac{{C_m(k) \,}}{{D_m(k) \,}}
\frac{{\epsilon_m J_m^2 \left( {knd} \right)}}{{2 R \ln a}}\,,
\end{equation}
where $D_m = (\ud A_m / \ud k)/R$ is given by
\begin{equation}
\begin{split}
D_m^{{\rm TM}} &  = J_m(knR) H_m^{\,\,''}(kR) - n^2 J_m^{\,\,''}(knR) H_m(kR), \\ 
D_m^{{\rm TE}} & = n \left[J_m(knR) H_m^{\,\,''}(kR) - J_m^{\,\,''}(knR) H_m(kR) \right] \\
& \qquad + (n^2  - 1) \,\, J_m^{\,\,'}(knR) H_m^{\,\,'}(kR) \nonumber,
\end{split}
\end{equation}
and $k$ is a resonance wavenumber of the unperturbed disk, with azimuthal modal index $m$.

%%%%%%%%%%%%%%%%%%%%%%%%%%%%%%%%%%%%%%%%%%%%%%%%%%%%

\section{Relation to a Finite Size Scatterer}

Before we investigate the solutions of Eq. (\ref{eq:quant_cond}) and the corresponding
resonance wave functions numerically, let us discuss the physical interpretation of the parameter
$a$. This parameter can be related to that of a small but finite size scatterer as long as 
this scatterer can be treated in the $s$-wave approximation. 
The Green's function of the system with a point scatterer follows from 
self-adjoint extension theory \cite{Zorbas1980,Shigehara1994}, and is given by
\begin{equation} \label{greenperturb}
G^a({\bf r},{\bf r_0},k) = G({\bf r},{\bf r_0},k) + \frac{G({\bf r},{\bf d},k)
{\cal D} G({\bf d},{\bf r_0},k)}{1 - {\cal D} G^{\mathrm{sc}}({\bf d},{\bf d},k)} \; .
\end{equation}
Here $G^{\mathrm{sc}}({\bf d},{\bf d},k)$ is the Green's function in (\ref{eq:G_is}) 
for both arguments less than $R$ and without the term $- \ui H_0(n k |{\bf r} - {\bf r}_0|)/4$
(i.e. without the free Green's function). The quantity ${\cal D}$ is the so-called diffraction
coefficient
\begin{equation} \label{diffcoeff}
{\cal D} = \frac{2 \pi}{\ui \pi/2 - \gamma - \log(n k a/2)} \; .
\end{equation}
One can easily check that the resonance wavenumbers in (\ref{eq:quant_cond}) coincide
with the poles of the Green's function (\ref{greenperturb}).

The form of the Green's function in (\ref{greenperturb}) is identical to that of a system
perturbed by a small $s$-wave scatterer within the Geometrical Theory of Diffraction
\cite{Keller1962,Vattay1994}. The optical theorem imposes a restriction on the diffraction
coefficient given by $|{\cal D}|^2 = - 4 \Im {\cal D}$ if $k \in \mathbb{R}$.
It is a consequence of the conservation of energy. The condition for ${\cal D}$ is equivalent to
$\Im {\cal D}^{-1} = 1/4$. In order to find a physical interpretation for the parameter $a$ one
has to compare the diffraction coefficient (\ref{diffcoeff}) with that of a small but finite
size scatterer in the $s$-wave approximation. In fact, our parameter $a$ was chosen in such
a way that the coefficient (\ref{diffcoeff}) agrees with that of a small $s$-wave scatterer
of radius $a$ with Dirichlet boundary conditions $\Psi=0$ along its circumference (see \cite{Rosenqvist1996}).

Let us now perturb the microdisk by a small hole of radius $b$ 
placed at a distance $d$ from the center, and filled with a material of refractive index $n_b$. 
The diffraction coefficient depends only on local properties near the perturbation
and can be obtained from the Green's function for a small circular disk of radius $b$
with refractive index $n_b$ embedded within material of refractive index $n$
(i.e. one can neglect the outside region with refractive index 1). But this Green's
function can be obtained in exactly the same manner as the Green's function in
Eq. (\ref{eq:G_is}). The only differences are the corresponding radius
and the refractive indices. In particular, the new coefficients are
\begin{equation} \nonumber
\begin{split}
{\cal A}_m^{\rm TM}& = n\,J_m(kn_bR)H_m^{\,\,\,'}(knR) - n_b\,J_m^{\,\,\,'}(kn_bR)H_m(nkR), \\
{\cal A}_m^{\rm TE}& = n_b\,J_m(kn_bR)H_m^{\,\,\,'}(knR) - n\,J_m^{\,\,\,'}(kn_bR)H_m(nkR), \\
{\cal B}_m^{\rm TM}& = n\,J_m(kn_bR)J_m^{\,\,\,'}(knR) - n_b\,J_m^{\,\,\,'}(kn_bR)J_m(nkR),  \\ 
{\cal B}_m^{\rm TE}& = n_b\,J_m(kn_bR)J_m^{\,\,\,'}(knR) - n\,J_m^{\,\,\,'}(kn_bR)J_m(nkR).
\end{split}
\end{equation}
We are interested in the Green's function outside the small disk. If that disk is
small enough to be treated in the $s$-wave approximation we need to keep only the
term $m=0$ in the sum over modal indices
\begin{align} \label{greenperturb2}
G^b({\bf r},{\bf r_0},k) & = -\frac{\ui}{4} H_0(k n |{\bf r} - {\bf r_0}|) 
\notag \\ & \qquad
+ \frac{\ui}{4} \frac{{\cal B}_0}{{\cal A}_0} H_0(k n |{\bf r}|) H_0(kn|{\bf r_0}|) \,.
\end{align}
This function has the form of (\ref{greenperturb}) with
$ G({\bf r},{\bf r_0},k) = - \ui H_0(k n |{\bf r} - {\bf r_0}|)/4$,
$ G^{\mathrm{sc}}({\bf r},{\bf r_0},k) = 0$, and we can read off the diffraction 
coefficient as ${\cal D} = - 4 \ui {\cal B}_0 / {\cal A}_0$. 
One can check that the optical theorem is satisfied. 
Using the asymptotic forms of Bessel and Hankel functions for small arguments, 
we obtain that ${\cal D}^{-1} \approx \ui/4 - [\pi b^2 k^2 (n_b^2 - n^2)]^{-1}$.
The comparison with the diffraction coefficient for 
a point scatterer (\ref{diffcoeff}) gives the condition on the parameter $a$
of the point scatterer for modelling the finite size scatterer
\begin{equation} \label{finite}
\ln \frac{n k a}{2} + \gamma \approx \frac{2}{b^2 k^2 (n_b^2 - n^2)} \; .
\end{equation}
In this equation $k$ is real, i.e. its imaginary part which is
small for resonances with high $Q$ factor is neglected.
Finally, we should note that the $s$-wave approximation itself is valid 
if the small but finite size scatterer is located not too close to the circumference of the disk, 
i.e. for $| n k (R-d)| \gg 1$, and if
\begin{equation}
\frac{{\cal A}_0}{{\cal B}_0} \ll \frac{{\cal A}_1}{{\cal B}_1} \; . 
\end{equation}

%%%%%%%%%%%%%%%%%%%%%%%%%%%%%%%%%%%%%%%%%%%

\section{Far Field Directivity}

In order to quantify the far-field behavior of the field $\Psi$ we consider
its asymptotics for $r\rightarrow\infty$ which has the form
\begin{equation} \label{ff}
\Psi\left( {\bf r}, k_{{\rm res}} \right) = \Psi\left(r,\varphi, k_{{\rm res}} \right) 
\propto \frac{{\exp (ik_{{\rm res}} r)}}{{\sqrt r }}f\left( \varphi  \right).
\end{equation}
Then, the directionality of the emission can be characterized
with the directivity $D$ which is defined as the ratio
of the power emitted into the main beam direction $\varphi_{\text{max}}$ to the total power
averaged over all possible directions,
\begin{equation}
\label{eq:directivity}
D = \frac{2\pi\, |f_{{\rm max}}(\varphi_{{\rm max}})|^2}  
{{\int\limits_0^{2\pi } {\left| {f( \varphi)} \right|^2 d\varphi}}}\,.
\end{equation}
Note that the resonances of the disk
without a scatterer have $D=1$ for $m=0$ and $D=2$ for $m\ne0$.

For the disk with a point scatterer, 
we investigate the level dynamics of the resonances as a function of the parameter $a$
by numerically solving Eq.~(\ref{eq:quant_cond}) for TM and TE modes, respectively.
As mentioned in Sec.~II and discussed in more detail in \cite{DettmannMorozovSieberWaalkens2008},
the perturbed resonances reduce to the unperturbed resonances (microdisk without scatterer)
in the limits $a\to 0$ and $a\to \infty$.  The wavenumbers of the perturbed disk thus evolve a long line segments in the complex wavenumber plane which start and end at unperturbed resonances when $a$ is varied from 0 to infinity. 
In our numerical procedure which consists of a Newton method to solve Eq.~(\ref{eq:quant_cond})  
we vary $a$ from  $10^{-30}$  to $10^{\,30}$ and use 
Eq.~(\ref{eq:approx_quant_cond}) to obtain starting points for the numerics in these limits. 
For each  resonance wavenumber found this way, the corresponding wavefunction is obtained from Eq.~(\ref{wave}).
In particular, in the outside region the wavefunction takes the form of (\ref{eq:outside_field}) from which we  can compute the directivity $D$. 

In fact, because of the relation between wavefunction and the Green's function (\ref{wave})
we can formally associate a directivity $D$ to any point in the complex $kR$ plane
(for a fixed choice of ${\bf d}$). This directivity has only a direct physical interpretation
if the $kR$ value corresponds to a resonance of the disk with
a point scatterer.  
For general $k$, $D(k)$  is rather a characterization of the Green's function (\ref{eq:G_is}).
It is helpful  to consider also this formal directivity because it enables one to clearly identify the regions in the 
$kR$ plane where a perturbation of the circular disk by a scatterer (at position ${\bf d}$)
should lead to highly directional modes.

\begin{figure}[htb]
\centerline{
\includegraphics[width=8.2cm]{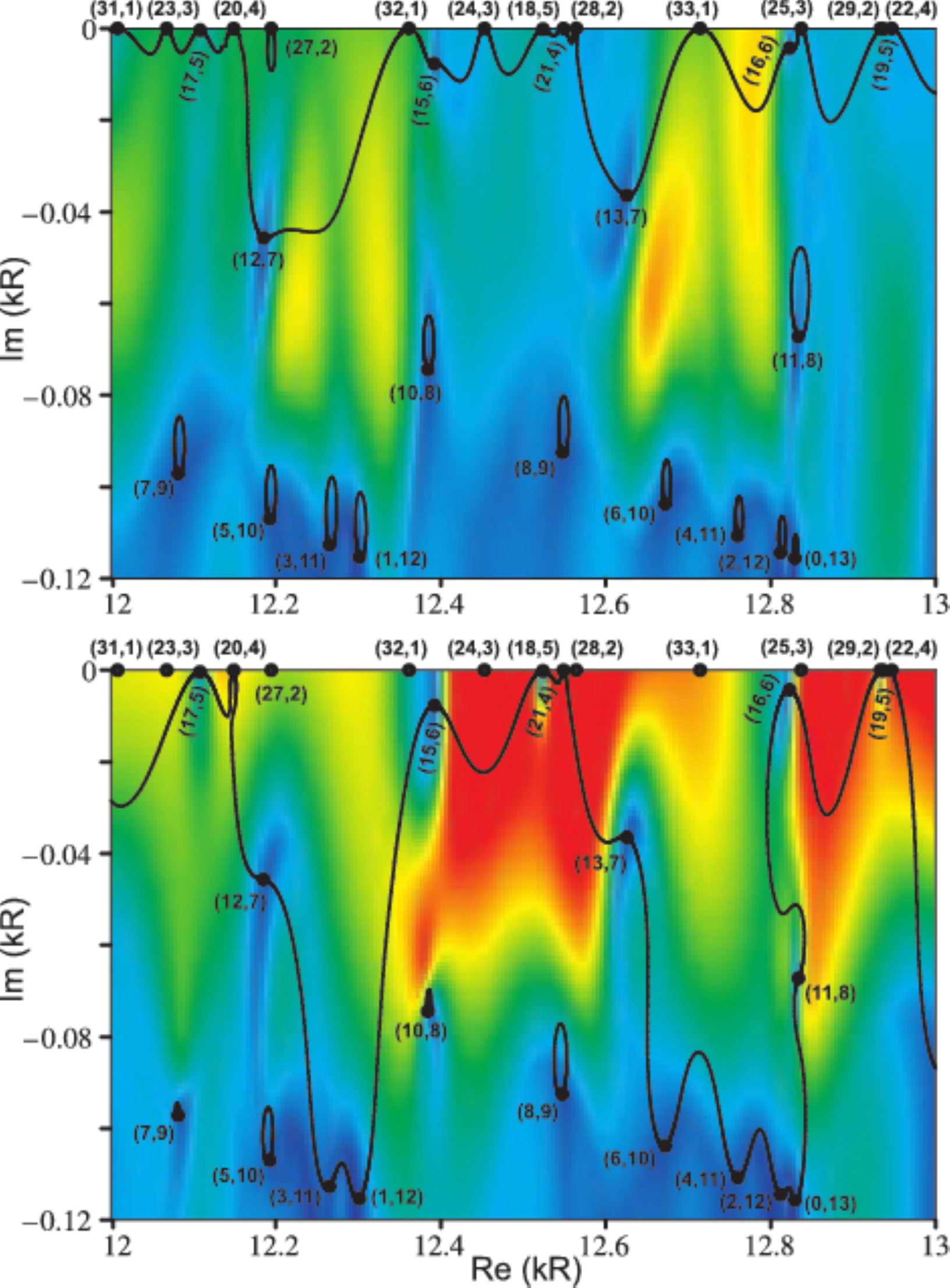}
}
\caption{(color online). Level dynamics (black curves) of the TM resonances
in the complex wave number plane for a dielectric disk with
$n=3.0$ and $R=1.0\,\mu$m and a point scatterer of varying
coupling parameter $a$. The solid circles mark the unperturbed
resonances with azimuthal and radial modal indices  $(m,q)$.
For the upper panel the scatterer is placed at distance $d=0.9$ $\mu\text{m}$
from the centre, for the lower panel $d=0.5$ $\mu$m. 
The color code in the $kR$ plane indicates the  directivity $D$
of the emission as explained in the text (blue marks small
values of $D$, red marks high values of $D$).} 
\label{Fig1}
\end{figure}
As a first example, we study the level dynamics of TM resonances in the range
$12 < {\rm Re}(kR) < 13$ for a disk of effective refractive index $n=3.0$
and radius $R=1\,\mu$m (these parameters are close to the ones in 
Ref.~\cite{Peter2005}) with a point scatterer placed at two different
distances ($d=0.9\,\mu$m and $d=0.5\,\mu$m) from the center of the disk.
The results are shown in Fig.~\ref{Fig1}.  The background color in Fig.~\ref{Fig1} 
indicates the directivity $D$ for any point of the $kR$ plane computed as mentioned above. Low values of $D$ correspond to the color blue, and high values of $D$ correspond to deep red.  Superimposing the curves of the wavenumber level dynamics we can immediately see in
which region of the $kR$ plane the highly directional modes are located. 

\begin{figure}[htb]
\centerline{
\includegraphics[width=8.2cm]{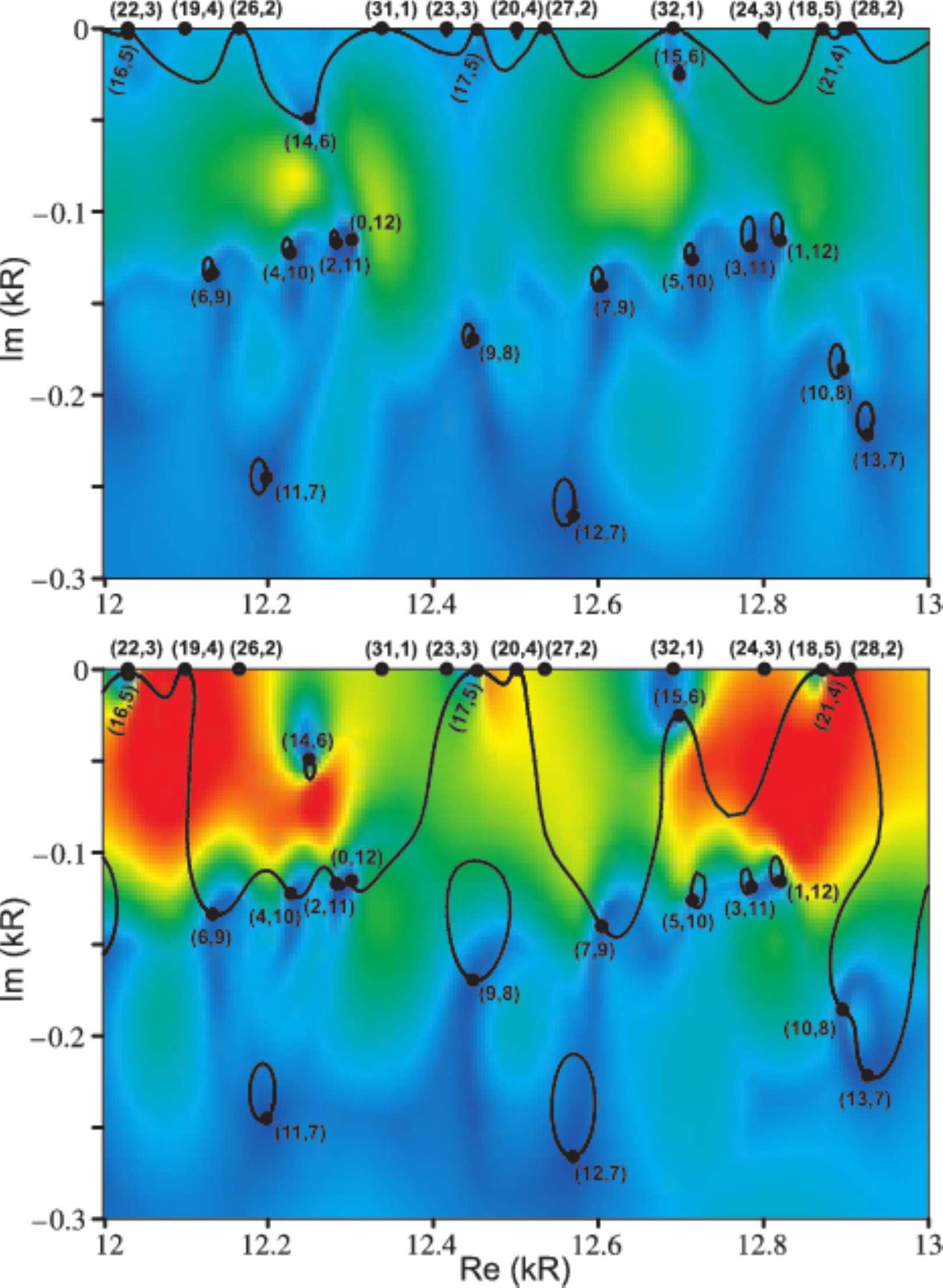}
}
\caption{(color online). Analogue of  Fig.~\ref{Fig1}  for TE resonances.
The parameters are the same as in Fig.~\ref{Fig1}.} 
\label{Fig2}
\end{figure}
Figure~\ref{Fig2} shows the corresponding plot for TE modes in the
same parameter range as in Figure \ref{Fig1}.  
Like in the case of TM polarization there are highly directional TE modes
for a wide range of $Q$-factors.
From the figures one sees that for both TM and TE resonant modes,
 placing  the point scatterer at  a distance  $d=0.5 \,\mu$m from the center
yields a better directivity than for $d=0.9 \,\mu$m. We will explain this observation 
in Sec.~\ref{geo}.
The directivity reaches values as high as $D \approx 15$
for some specific TM modes and $D \approx 13$ for some specific TE modes.

\begin{figure}[htb]
\centerline{
\includegraphics[width=8.2cm]{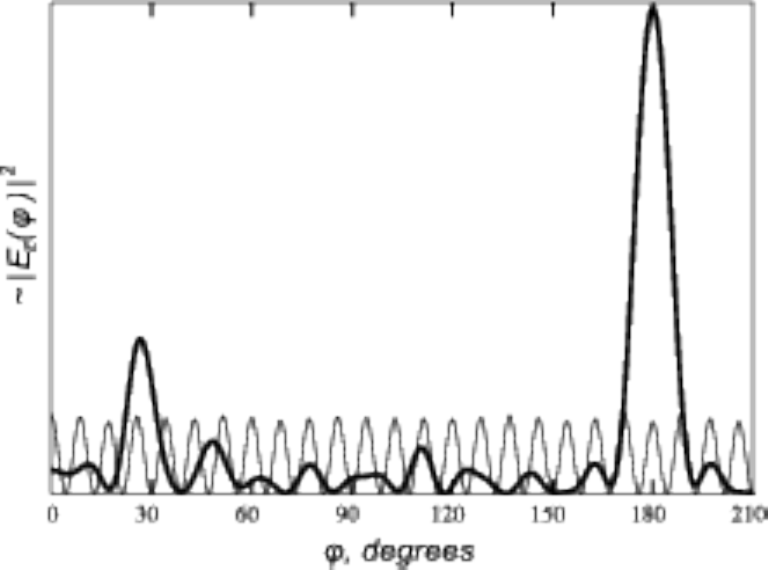}
}
\caption{The far-field intensities  
of the unperturbed TM resonance with
modal indices $(21,4)$ (thin curve), and of the
highly directional TM resonance (thick curve) that
is obtained from the unperturbed mode by perturbing
a dielectric disk with $n=3.0$ and $R=1.0\,\mu$m
with a point scatterer of strength $a=10^{-6}$
placed at distance $d=0.5\,\mu$m from the disk center.}
\label{Fig3}
\end{figure}

\begin{figure}[htb]
\centerline{
\includegraphics[width=8.2cm]{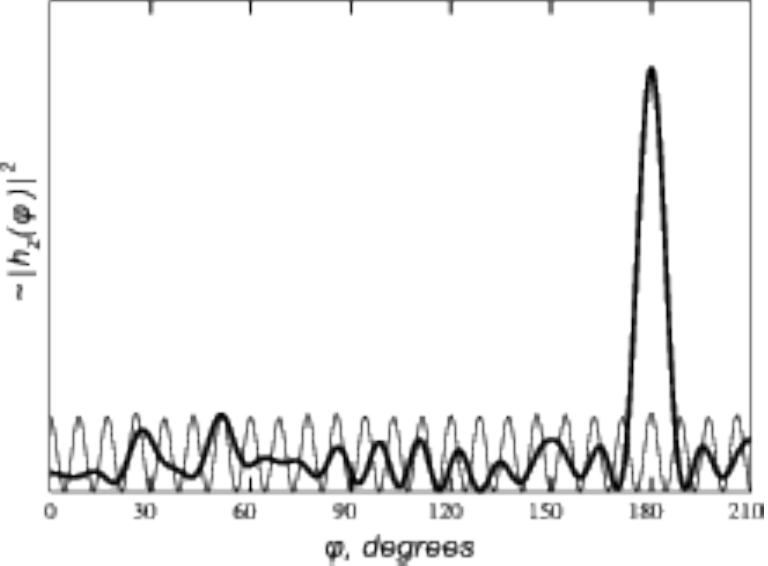}
}
\caption{The field intensity in 
the far-field of the unperturbed TE resonance with
modal indices $(21,4)$ (thin curve), and of the
highly directional TE resonance (thick curve) that
is obtained from the unperturbed mode by perturbing
a dielectric disk with $n=3.0$ and $R=1.0\,\mu$m
with a point scatterer of strength $a=10^{-6}$
placed at distance $d=0.5\,\mu$m from the disk center.}
\label{Fig4}
\end{figure}
In Fig.~\ref{Fig3} we show the field intensity at the distance
$r=500~\mu$m (far-field region) for a highly directional
TM resonance mode which is obtained by a small perturbation of the
unperturbed TM resonance mode with modal indices $(21,4)$. It has
directivity ($D \cong 13$) and complex wavenumber
$kR=12.54929-{\ui}\,0.000045$, and is compared to the unperturbed
mode with directivity $D = 2$ and complex wavenumber
$kR=12.54876-{\ui}\,0.000001$. The highly directional mode is
obtained from the unperturbed mode if a very weak point scatterer
of strength $a \cong 10^{-6}$ is placed at a distance $d=0.5\,\mu$m
from the centre of the disk. Indeed, we expect this from Fig.~\ref{Fig1},
because the unperturbed mode lies in a highly red region in the $kR$
plane where a small perturbation should lead to a highly directional mode.
According to Eq. (\ref{finite}) the perturbation is comparable to that
of a finite size scatterer of radius $b\approx 0.01 \mu m$ and refractive index $n_b=1$.

In Fig.~\ref{Fig4} we show the far field intensity of the analogous
TE mode. It is again obtained by a perturbation of the mode $(21,4)$
of the circular disk by placing a point scatterer of strength
$a \cong 10^{-6}$ at distance $d=0.5\,\mu$m from the centre of the
disk. For this polarization the unperturbed mode $(21,4)$ is located
at $kR=12.90089-{\ui}\,0.000001$ in the complex wave number plane
while the perturbed mode is located at $kR=12.90187-{\ui}\,0.000087$.
Their directivities are $D = 2$ and $D \cong 11$ respectively.
Both of the above highly directional modes are good candidates for lasing 
since their $Q$-factors are large. In fact, $Q \approx 1.4 \cdot 10^{5}$
for the highly directional TM resonance and
$Q \approx 7.4 \cdot 10^{4}$ for the highly directional TE resonance.
Moreover, they have unidirectional emission at an angular direction of
180 degrees.

\begin{figure}[htb]
\centerline{
\includegraphics[width=8.2cm]{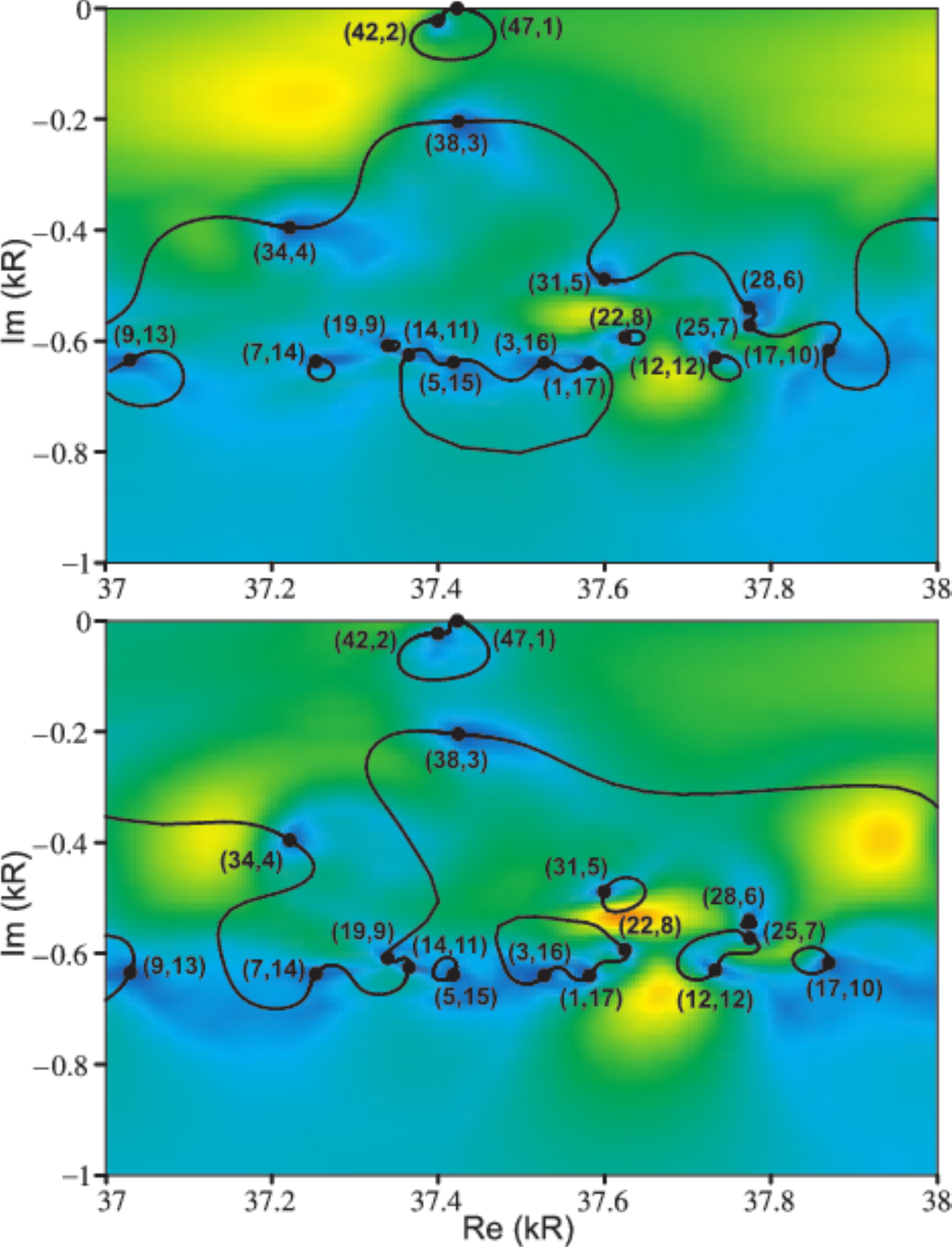}
}
\caption{(color online). Level dynamics of the TM resonances 
in the complex wave number plane 
for a dielectric disk with $n=1.4$ and $R=10.51$~mm 
and a point scatterer of varying coupling parameter $a$. 
The solid circles mark the unperturbed resonances with 
azimuthal and radial modal indices  $(m,q)$.
For the upper panel the scatterer is placed at distance $d=9.585$~mm from the centre, 
for the lower panel $d=9.385$~mm. 
As before, the color code indicates the  directivity $D$ of the emission.} 
\label{Fig5}
\end{figure}
\begin{figure}[htb]
\centerline{
\includegraphics[width=8.2cm]{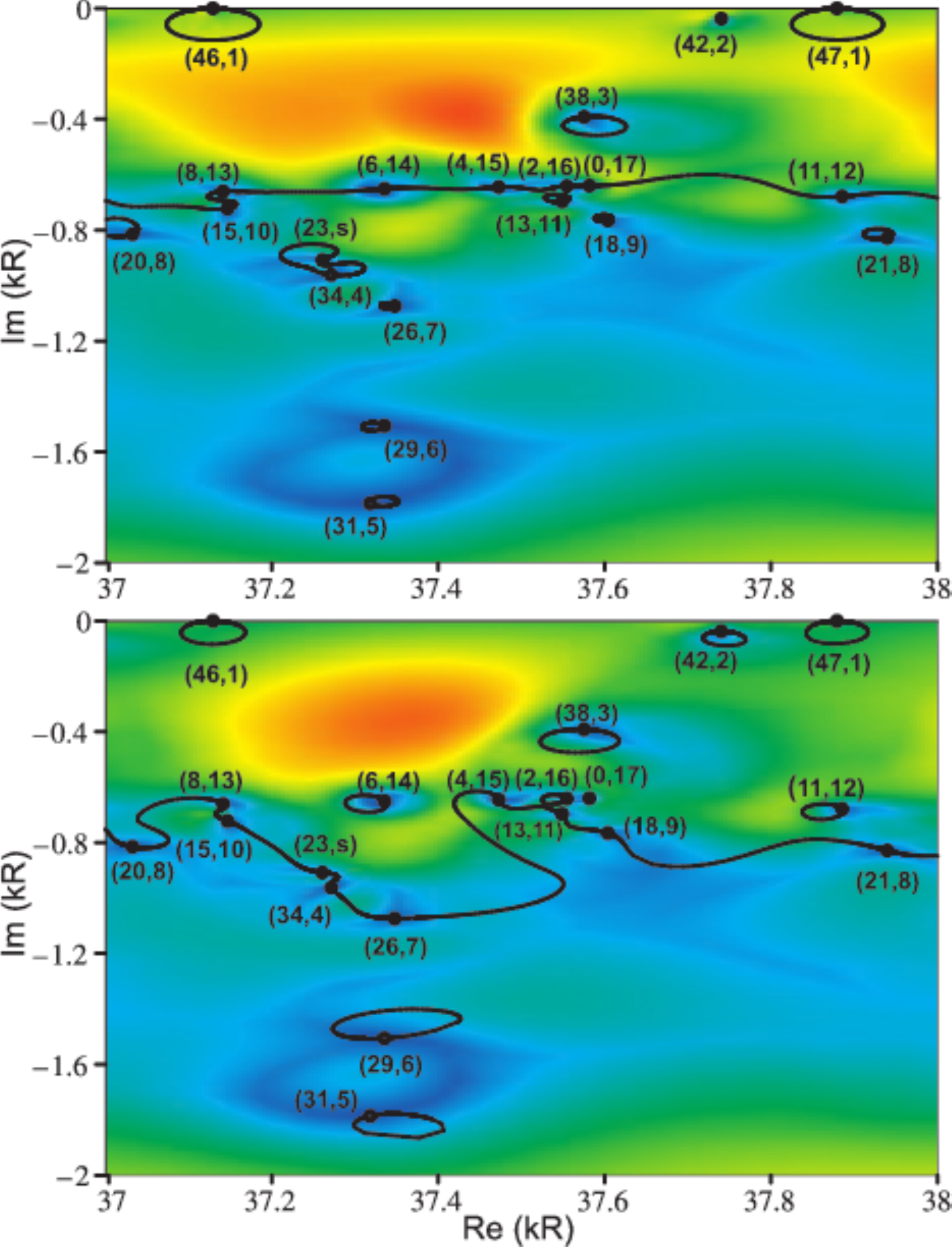}
}
\caption{(color online). Analogue of Fig.~\ref{Fig5} for  TE resonances.
The parameters are the same as in Fig.~\ref{Fig5}. The label
$(23,s)$ indicates a shape resonance with azimuthal model index $m=23$.}
\label{Fig6}
\end{figure}

As a second example, we study the level dynamics of TM and TE resonances 
in the range $37 < {\rm Re}(kR) < 38$
for a disk of effective refractive index $n=1.4$ and radius $R=10.51$~mm
(these parameters are close to the ones of the experimental setup of Schwefel and Preu,
\cite{Schwefel2008}) with a point scatterer placed at two close distances $d=9.585$~mm
and $d=9.385$~mm from the centre of the disk, see figs.~\ref{Fig5} and \ref{Fig6}.
From these two figures one can see that even a relatively small change in the position
of a point scatterer can lead to a significant change in the level dynamics.  
The directivity of some perturbed TM modes reaches values as high as 
$D \cong 10$ for a point scatterer located at $d=9.385$~mm, 
see the lower panel of Fig.~\ref{Fig5}, 
and the directivity of some TE modes reaches values as high as
$D \cong 8$ for a point scatterer located at $d=9.585$~mm, 
see the upper panel of Fig.~\ref{Fig6}. In general, however, the
directivity is lower than for $n=3$ in Figs.~\ref{Fig1} and \ref{Fig2}.
This will be explained in the next section.

Another interesting feature is the appearance 
of the unperturbed shape (external) TE resonance denoted by $(23,s)$  in Fig.~\ref{Fig6}
which is located in the region of the complex wave number plane 
where one would expect to have internal resonances only.
This is related to a peculiar behaviour of some 
of the unperturbed external TE resonances that was observed in 
Ref.~\cite{DettmannMorozovSieberWaalkens2009}. 
For large refractive indices $n$, the external resonances are located much deeper
in the complex wavenumber plane than the internal resonance. This clear  separation
by the magnitude of the imaginary part  of the wavenumbers ceases to exist for  small
$n$ in the case of TE modes where some external resonances mix with the internal
resonances. The wavefunctions
of such external resonances acquire similar features as the wavefunctions of internal
resonances. In particular they can spoil the interpretation of the radial modal index
$q$ as the number of peaks of the wavefunction in the radial direction
(see \cite{DettmannMorozovSieberWaalkens2009}) for internal resonances
with the same azimuthal modal index. 
Furthermore for the disk with a point scatterer, the external resonance
(23, s) in Fig. 6 illustrates the fact that unperturbed external
resonances also serve as starting and end points of the line segments
that result from the level dynamics of perturbed resonances in the complex
wavenumber plane upon varying $a$ from 0 to $\infty$.

\begin{figure}[htb]
\centerline{
\includegraphics[width=8.2cm]{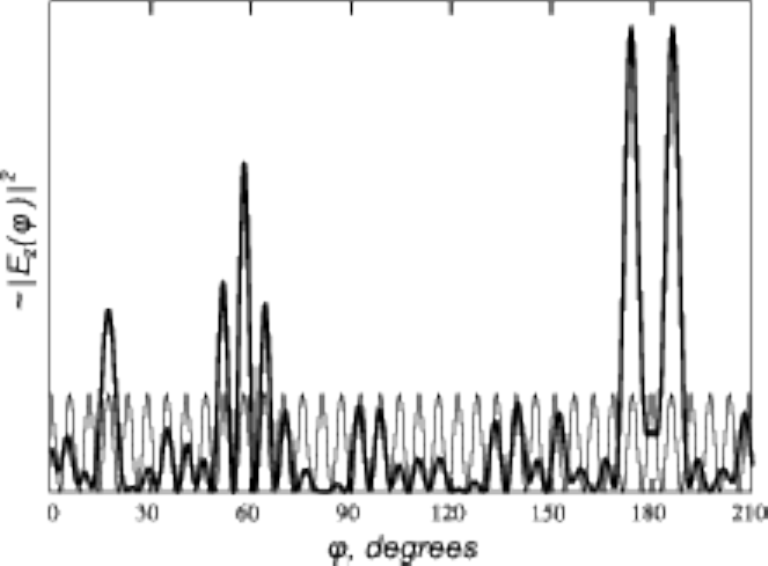}
}
\caption{The far-field intensities ($r=5000$~mm) 
of the unperturbed TM resonance 
with $kR=37.599462-{\ui}\,0.488553$, $m=31$, $q=5$ (thin curve),
and of the highly directional perturbed TM resonance
with $kR=37.621007-{\ui}\,0.523645$, $d=9.385$~mm, 
$a \protect\cong 0.07$ (thick curve)
in a dielectric disk with $n=1.4$ and $R=10.51$~mm.}
\label{Fig7}
\end{figure}
\begin{figure}[htb]
\centerline{
\includegraphics[width=8.2cm]{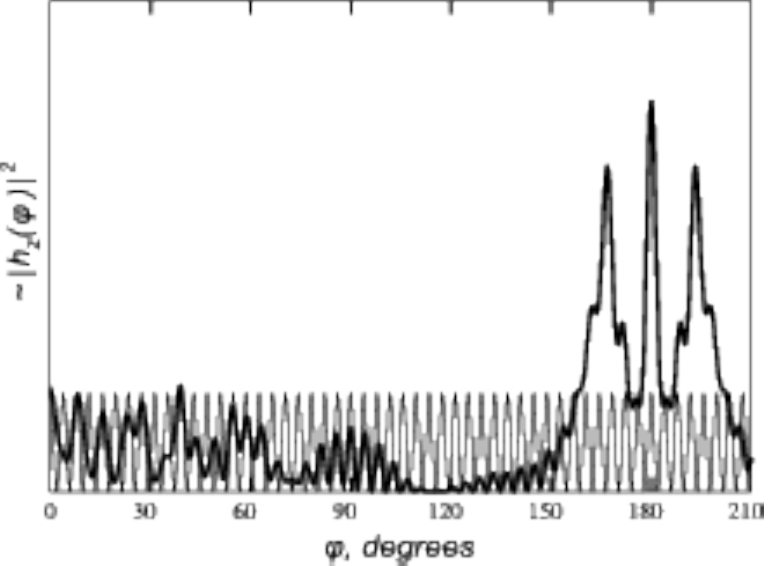}
}
\caption{The far-field intensities
 ($r=5000$~mm) 
of the unperturbed TE resonance 
with $kR=37.129055-{\ui}\,0.000177$, $m=46$, $q=1$ (thin curve),
and of the directional perturbed TE resonance
with $kR=37.142373-{\ui}\,0.001832$, $d=9.585$~mm,  $a \protect\cong 4 \cdot 10^{-5}$  (thick curve)
in a dielectric disk with $n=1.4$ and $R=10.51$~mm.}
\label{Fig8}
\end{figure}
In Fig.~\ref{Fig7} we show the field intensities at the distance
$r=5000$~mm (far-field region) for the unperturbed resonant TM mode 
with modal indices $(31,5)$, complex wavenumber $kR=37.599462-{\ui}\,0.488553$,
and directivity $D = 2$ as well as for the highly directional ($D \cong 10$)
TM resonant mode with complex wavenumber $kR=37.621007-{\ui}\,0.523645$.
The unperturbed mode $(31,5)$ transforms to the highly directional mode
with the above complex wavenumber if we place a point scatterer of
strength $a \cong 0.07$ at distance $d=9.385$~mm from the centre of
the disk. We note that despite of the high directivity this mode
is not suitable for lasing because it has only a small $Q$ factor of about $36$.

In Fig.~\ref{Fig8} we show the field intensities at the distance $r=5000$~mm
(far-field region) for the unperturbed resonant TE mode with modal indices
$(46,1)$, complex wavenumber $kR=37.129055-{\ui}\,0.000177$, and directivity
$D = 2$ as well as for the directional ($D \cong 6$) TE resonant mode
with complex wavenumber $kR=37.142373-{\ui}\,0.001832$.
The unperturbed mode $(46,1)$ transforms to the directional mode
with the above complex wavenumber if we place
a point scatterer of strength $a \cong 4 \cdot 10^{-5} $ at distance $d=9.585$~mm
from the centre of the disk.
This mode has $Q \cong 10^{4}$ and, therefore, is suitable for lasing.

%%%%%%%%%%%%%%%%%%%%%%%%%%%%%%%%%%%%%%%%%%%%%%%%%%%%%%%%%%%%%%%%%%%

\section{Directivity and Geometric Optics} \label{geo}

\begin{figure}[htb]
\includegraphics[width=6.5cm]{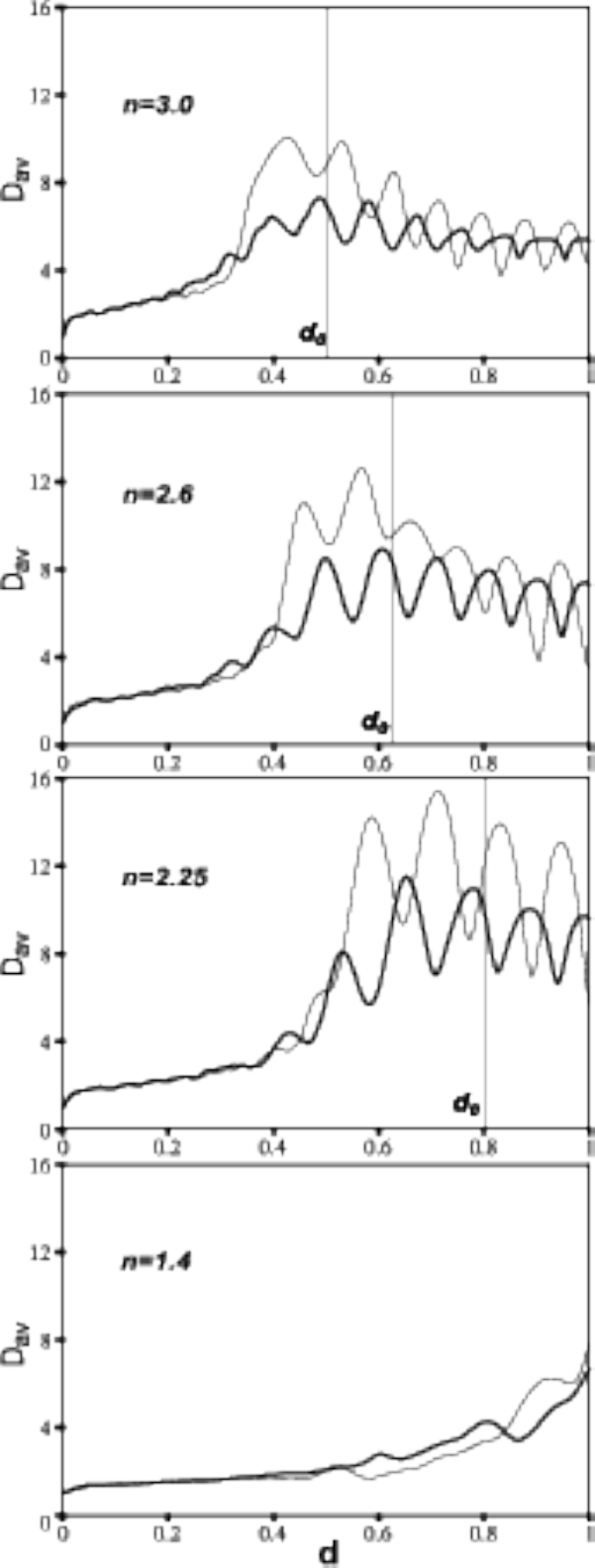}
\caption{The average directivity, $D_{{\rm av}}$, of TM (thick curves) and TE (thin curves)
polarized light in the range $12 < {\rm Re}(kR) < 13$, $-0.1 < {\rm Im}(kR) < 0.0$ for each position, $d$,
of a point scatterer in microdisks of radius $R=1\,\mu$m 
and various effective refractive indices as shown.}  
\label{Fig9}
\end{figure}

\begin{figure}[htb]
\includegraphics[width=6.5cm]{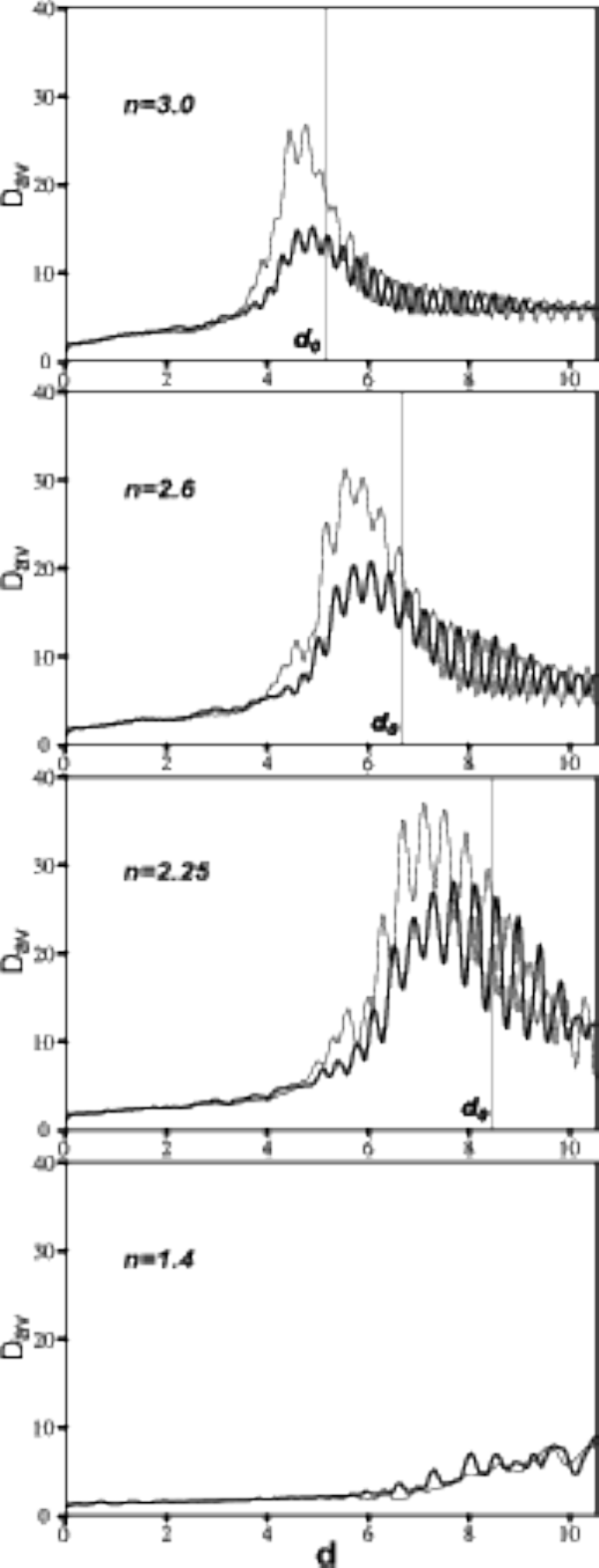}
\caption{The average directivity, $D_{{\rm av}}$, of TM (thick curves) and TE (thin curves)
polarized light in the range $37 < {\rm Re}(kR) < 38$, $-1 < {\rm Im}(kR) < 0.0$ for each position, $d$,
of a point scatterer in microdisks of radius $R=10.51\,\mu$m 
and various effective refractive indices as shown.}  
\label{Fig10}
\end{figure}

To systematically study the appearance of highly directional modes,
we calculate the average directivity for a  region in the complex
wavenumber plane as a function of the 
distance, $d$, of the point scatterer from the center. 
To this end we define the average directivity
\begin{equation}
\label{eq:directivity_average}
D_{{\rm av}}  = \frac{1}{\Delta k_r \Delta k_i} \int_{k_r^-}^{k_r^+}
\text{d}k_r   \int_{k_i^-}^{k_i^+}  \text{d}k_i\,   {D\left( k \right)}  \, ,
\end{equation}
where $[k^-_r,k^+_r]\times[k^-_i,k^+_i]$ is a rectangular region in the
complex wavenumber plane of side lengths $\Delta k_r = k^+_r-k^-_r$ and
$\Delta k_i = k^+_i-k^-_i$ .
Note that the integration is over all $k$ in the rectangular region where
$D\left( k \right)$ is formally defined using formulas
 (\ref{eq:outside_field}), (\ref{ff}), and (\ref{eq:directivity}), for
general $k$ rather than just for the resonant  $k_{\rm res}$ (see Sec.~IV).
In Figs.~\ref{Fig9} and \ref{Fig10}.
we  show $D_{{\rm av}}$ computed for the region
$12 < {\rm Re}(kR) < 13$, $-0.1 < {\rm Im}(kR) < 0$
as a function of $d$ 
for four microdisks of radius $R=1\,\mu$m and
effective refractive indices of 
$n=3.0$, $n=2.6$, $n=2.25$, and $n=1.4$,
and for the region $37 < {\rm Re}(kR) < 38$,
$-1 < {\rm Im}(kR) < 0$ for four disks of radius $R=10.51\,$mm 
with the same effective refractive indices, respectively.
Remarkably, 
a rough approximation of the values for $d$ which lead to high  $D_{{\rm av}}$ 
can be found from geometric optics. 
To this end
let us consider parallel rays that come in from infinity and enter
a dielectric disk of radius $R$ and effective refractive index $n$.
There is one ray which goes through the center of the disk, the
central ray. The rays that are infinitesimally close to this
central ray will cross the central ray at the point with
distance
\begin{equation}
\label{eq:geom_opt}
d_o=\frac{R}{n-1}
\end{equation}
to the center of the disk located on the opposite side of the center of the disk. 
So, conversely,  putting a point scatterer at this focal point leads to a strongly directional light emission in the 
an angular direction of 180$^\circ$ which also agrees with the observation in the previous section.

The value of $d_o$ is indicated by the vertical lines in Figs.~\ref{Fig9}
and Fig.~\ref{Fig10}. One can see that in most cases $d_o$ is close to
the optimal value range for the point scatterer positions in the figures.
Taking into account the finite size of the disk, we note that formula
(\ref{eq:geom_opt}) is valid only for refractive indices greater than $2$.
For smaller refractive indexes the optimal position should be 
as close as possible to the boundary of the disk.

%%%%%%%%%%%%%%%%%%%%%%%%%%%%%%%%%%%%%%%%%%

\section{Exceptional points}

The line segments that connect unperturbed resonances as the parameter
$a$ varies from zero to infinity can change considerably if the distance, $d$,
of the point scatterer to the center is changed.
This can be seen already in Figures~\ref{Fig5} and \ref{Fig6} which are for
two different but close values of
$d$. The connections between the unperturbed resonances are very
different there, i.e. the line segments connect 
different unperturbed resonances if $d$ is varied only slightly.
In the present section we want to look at this in more detail. 

\begin{figure}[htb]
\centerline{
\includegraphics[width=8.2cm]{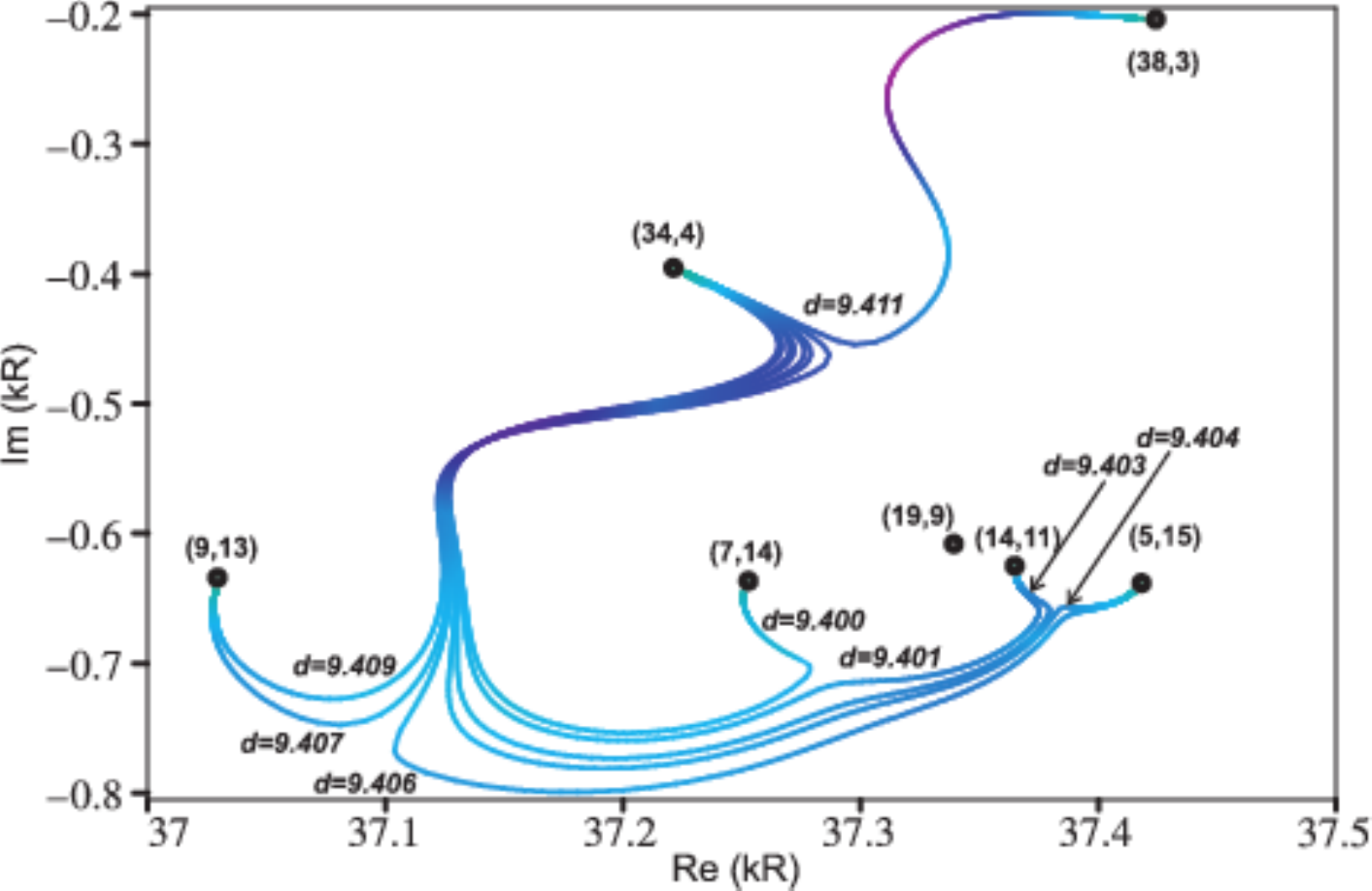}
}
\caption{(color online). Level dynamics of the TM resonances in the
complex wave number plane for a dielectric disk with $n=1.4$ and $R=10.51$mm 
and a point scatterer of varying coupling parameter $a$ for several
positions $d$ of the scatterer. All start from the unperturbed 
resonance $(34,4)$. The solid circles mark the unperturbed resonances with 
modal indices  $(m,q)$. The color code indicates the  directivity $D$
of the emission (green marks small values of  $D$, red marks high values of $D$).} 
\label{Fig11}
\end{figure}

As a first example we investigate the perturbation of one
particular resonance of the microdisc. We choose the TM resonance with
modal indices $(34,4)$ of the dielectric disk with $n=1.4$
and $R=10.51$. We then place a point scatterer at 
distance $d$ from the center of the disk. For different values
of $d$ we vary the strength of the scatterer $a$ from $0$
to infinity. This yields a family of line segments in the complex wavenumber
plane which all start at the unperturbed resonance $(34,4)$ but,
depending on $d$, end  at different  unperturbed resonances. 
This is shown in Fig.~\ref{Fig11}.
It is remarkable that the connections between the
different unperturbed resonances depend so sensitively
on the value of $d$. In the small range from $d=9.400$ to $d=9.411$
shown in  Fig.~\ref{Fig11} the unperturbed resonance $(34,4)$ is
connected to five different unperturbed resonances.

\begin{figure}[htb]
\centerline{
\includegraphics[width=8.2cm]{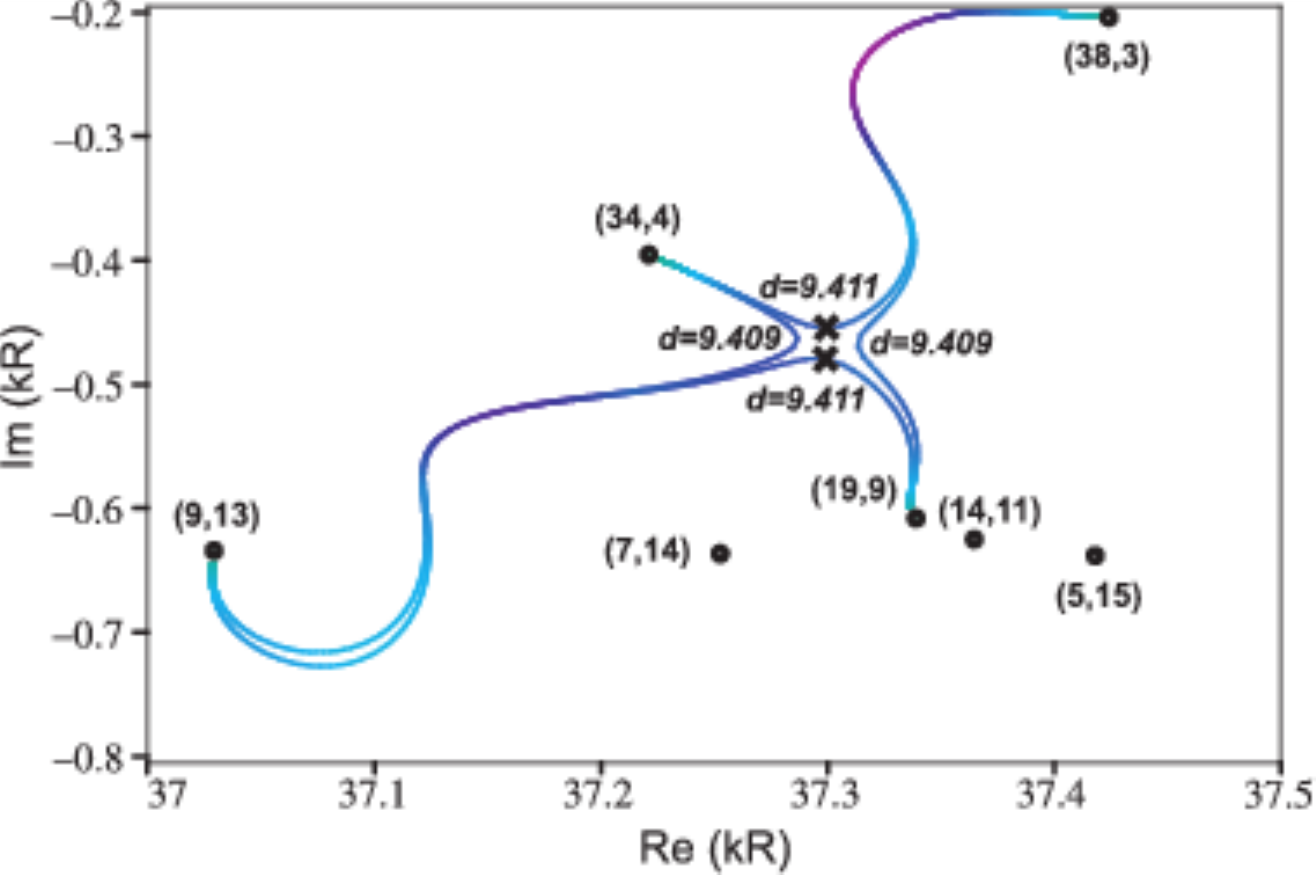}
}
\caption{(color online). Level dynamics of the TM resonances in the
complex wave number plane for a dielectric disk with $n=1.4$ and $R=10.51$mm 
and a point scatterer of varying coupling parameter $a$ for two
positions $d$ of the scatterer. The solid circles mark the unperturbed
resonances with azimuthal and radial modal indices  $(m,q)$.
As the position of the scatterer changes from $d=9.409$ to $d=9.411$
the connections of the unperturbed resonances change.} 
\label{Fig12}
\end{figure}

Next we investigate how the connections are rearranged. 
The connections between the unperturbed
resonances can only change upon variation of $d$ if for one value of $d$ different line segments intersect
tangentially   at  a point in the complex $kR$-plane. This point corresponds to a degenerate resonance.
This mechanism is illustrated in Fig.~\ref{Fig12}. For $d=9.409$
the resonance $(34,4)$ is connected to the resonance $(9,13)$, while
the resonance $(19,9)$ is connected to $(38,3)$. For the value $d=9.411$ the
connections have changed. The resonance $(34,4)$ is now connected
to $(38,3)$, while $(19,9)$ is connected to $(9,13)$. Although
not shown in the figure, there is a value of $d$ for which two line segments touch each other at a point at which two perturbed resonances
coallesce and become degenerate.

\begin{figure}[htb]
\centerline{
\includegraphics[width=8.2cm]{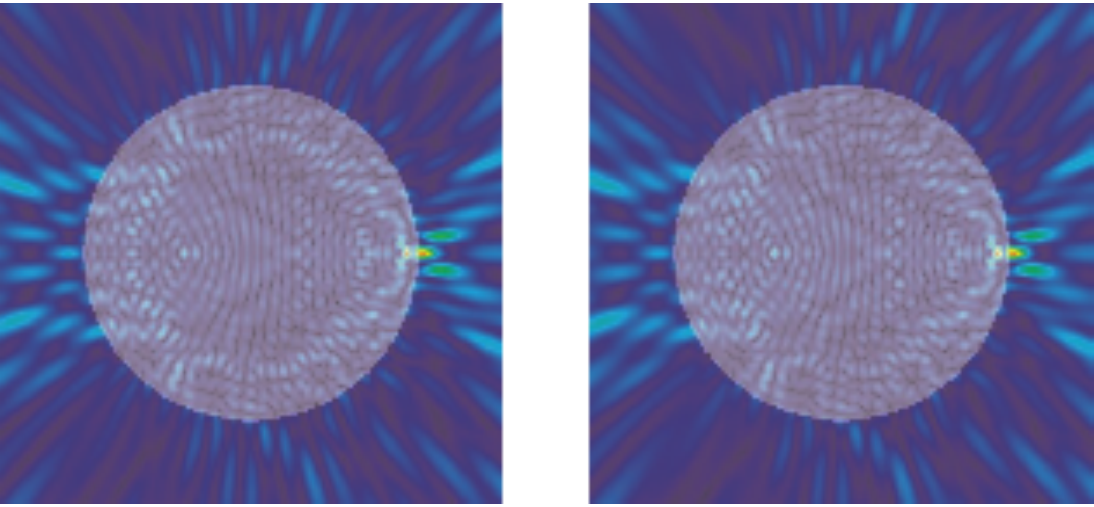}
}
\caption{(color online). Two almost degenerate TM resonance states for a
dielectric disk with $n=1.4$ and $R=10.51$mm. The corresponding wavenumbers
are shown by crosses in Fig. \protect\ref{Fig12}.} 
\label{Fig13}
\end{figure}
For an open system like the dielectric disk degeneracies are generically
of a special type which are called exceptional points
\cite{Heiss2000,Kato1966}. Exceptional points can be
observed if at least two real-valued parameters of a non-hermitian
operator are varied. 
In our case the non-hermitian operator is the differential operator acting on $\Psi$ in 
Eq.~(\ref{eq:Psi}) which is non-hermitian due to the outgoing boundary condition.  
The two real parameters are the parameters $a$ and $d$. 
At an exceptional point two (or more) eigenvalues of the non-hermitian
operator coallesce, and the corresponding eigenstates become 
identical. The pair of eigenstates which becomes degenerate at an exceptional point also show a characteristic behavior 
if the two parameters are changed along a closed loop about the  exceptional points. 
Smoothly following the pair of eigenstates from a starting point of the loop upon one full traversal of the loop 
yields a pair of  eigenstates in which the eigenstates started with are swapped and in which one member of the pair has a reversed sign. 
Exceptional points have recently 
received a lot of attention. 
For an example in the field of microlasers and more 
references see Ref.~\cite{WiersigKimHentschel2008}.

\begin{figure}[htb]
\centerline{
\includegraphics[width=8.2cm]{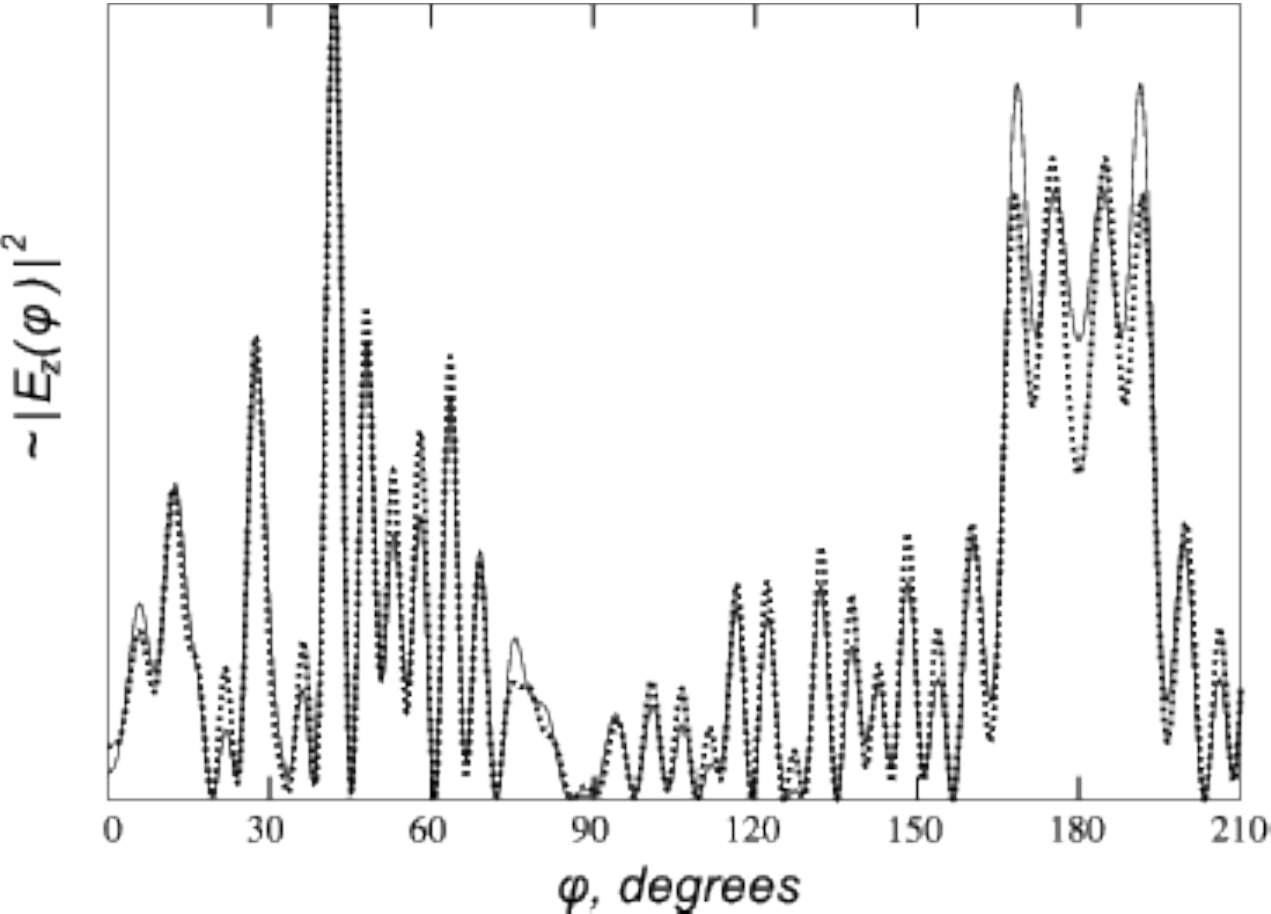}
}
\caption{Far field of the two resonance states shown by
crosses in Fig. \protect\ref{Fig12}.} 
\label{Fig14}
\end{figure}

To illustrate the coalescence of resonance states that is typical
for exceptional points we show in Fig.~\ref{Fig13} the fields of the
two almost degenerate resonances marked in Fig.~\ref{Fig12}. 
Fig.~\ref{Fig14} shows the corresponding far-field behaviour.

The numerical results in this section indicate that exceptional
points are quite common for dielectric disks with point scatterers.
They control the connections between the unperturbed resonances
and can be easily found by noting when these connections change
upon varying the position ${\bf d}$ of the scatterer.

%%%%%%%%%%%%%%%%%%%%%%%%%%%%%%%%%%%%%%%%%%%%%%

\section{Conclusion}

We have shown that perturbations of a dielectric disk by a point
scatterer can lead to highly directional resonance modes with 
large Q-factors. This is demonstrated in particular by the
unidirectional modes in Figs.~\ref{Fig3} and \ref{Fig4}.
To obtain modes with these properties is one of the main goals
in the design of dielectric microcavities. 

The system studied has the advantage that it is relatively simple and
 can be treated to a large extent analytically by a Green's function method.
This allows for a systematic investigation of the system over a large parameter
range with only moderate numerical effort. We found that several
numerical results can be understood with the help of a
simple geometrical optics model. This model helps to find the 
optimal position of the scatterer, it explains why the directivity
is in general higher for refractive index $n=3$ than for $n=1.4$,
and also why the emission occurs predominantly in a  direction
opposite of the position of the point scatterer. 
It also suggests future investigation of an elliptical microcavity
with a scatterer at one of the foci and refractive index inverse to
the eccentricity, as in that case focusing in geometric optics is exact,
i.e. the paraxial approximation is not required \cite{BNBBSA2004}.

The Green's function method also allows one to associate a directivity
with different regions of the complex $kR$ plane. This is very
useful because it indicates in which regions of the $kR$ plane
one can expect highly directional modes if one perturbs the
dielectric disk by the scatterer. It would be helpful
to find a semiclassical explanation for the
dependence of the directivity on the wavenumber $k$.

Most interesting for applications we also discussed how the system
studied can be realized physically by a small but finite sized scatterer.
This connection can be made as long as the scatterer can be treated in
the $s$-wave approximation, and is found to be valid for examples with
high Q-factor and directivity. An important open question is how the
directivity and $Q$-factor depend on the size and shape of a larger
scatterer when corrections to the $s$-wave approximation are taken
into account.

\section*{Acknowledgements}

This work was supported by the EPSRC under grant number EP/C515137/1.

%\bibliographystyle{apsrev}   %>>>> makes bibtex use the correct BibTeX style file (bst file)
%\bibliography{TmTeDisk}   %>>>> bibliography data from microlaser.bib

\end{document}